\documentclass[fleqn,twoside]{article}
\usepackage{espcrc2}
\usepackage{epsfig}

\usepackage{graphicx}

\title{Progress in lattice algorithms}

\author{Mike Peardon\address[TCD]{School of Mathematics, Trinity College,
        Dublin, Dublin 2, Ireland.}
        }
       
\begin{document}

\begin{abstract}
The development of Monte Carlo algorithms for generating gauge field 
configurations with dynamical fermions and methods for extracting the most 
information from ensembles are summarised. 
\vspace{1pc}
\end{abstract}

\maketitle

\section{INTRODUCTION}

  This conference illustrated that most large collaborations actively 
investigating QCD physics are performing simulations of the theory with 
dynamical quark fields. Dynamical staggered fermion simulations have been 
pursued for many years now, however it is clear that the community is
now also firmly in the era of full Wilson fermion studies.  These calculations 
will address one of the biggest systematic error in many non-perturbative QCD 
predictions from the lattice; use of the ``quenched approximation''.
At present, most simulations are not being performed at realistic values of the 
quark masses, or with the correct number of flavours of light fermions. 
A technically challenging extrapolation of data to the up and down quark masses 
must subsequently be performed. 

  At the conference, statements were given \cite{rdkpanel} by the biggest 
collaborations.  Estimated costs for reproducing the state-of-the-art in 
quenched spectrum data in the full theory were in the range 10-100 Tflop-years.
In this review, efforts to develop better algorithms to simulate closer to 
the physical parameters of QCD are discussed.

\section{ALGORITHMS FOR WILSON FERMION SIMULATIONS}

  In this section, the main features of the 
techniques for simulating dynamical Wilson fermions are summarised. 
The simulations of dynamical staggered fermions 
are more mature; the principle difficulty at present is 
one of controlling the choice of discretisation \cite{Toussaint:2001zc}.
 The new challenge is to find 
efficient methods for including recently proposed improved discretisations. 
Developing algorithms incorporating ``fat-link'' gauge fields, 
designed to enhance the flavour symmetry of the action are presented later.
It is worth noting that the exact algorithms developed 
for odd-flavour simulations with Wilson fermions are being applied to staggered 
fermion simulations.

The Hybrid Monte Carlo (HMC) algorithm \cite{Duane:1987de} is still the 
work-horse powering most large Wilson fermion simulations. 
Over the past few years 
there have been further advances in our understanding of the issue of finite
precision arithmetic and the reversibility and stability of the molecular 
dynamics (MD)
integrator component. HMC is an exact algorithm if the MD
integrator is reversible and area preserving. 
It has long been understood that the Hamiltonian evolution of QCD is chaotic,
and the Liapunov exponents were studied in detail 
\cite{Edwards:1996vs,Liu:1997fs}. 
More recently, the stability of the MD leap-frog integrator 
\cite{Edwards:1996vs} has been 
investigated in detail by the UKQCD collaboration \cite{Joo:2000dh}. Their 
findings were presented to the conference by Jo\'o \cite{Joo:2001ky}. 
Analysis of a leap-frog integrator of the simple 
harmonic oscillator reveals the solution diverges once the step-size exceeds a
critical value. The problem is exacerbated for higher-order integrators. This 
same behaviour is observed in QCD, and the critical step-size is seen to fall
as the quark mass, $m_q \rightarrow 0$. The suggestion is that the problem of 
simulating light Wilson fermions with HMC will amount to managing this 
stability problem. 

  There has been progress in implementations of the local boson algorithm (LBA)
\cite{Luscher:1993xx}.
This algorithm relies on first constructing a polynomial 
approximation to the inverse of the fermion matrix, then expressing the
polynomial as a product of roots. The determinant is 
expressed as a local bosonic partition function involving a large number (the
degree of the polynomial) of auxiliary fields. The beauty of the method is that
a local change in the gauge fields leads to a local computation for the change 
in the action.
Unfortunately, the simplest implementations of this method suffer from long 
autocorrelations \cite{Jansen:1996xp}; a large number of auxiliary fields are
coupled to the gauge fields, and so only small changes in these degrees of 
freedom can be made in
each sweep. The problem can be circumvented by using lower-order polynomials, 
and correcting for any discrepancy between the approximation and the true 
inverse using a noisy accept-reject step \cite{Borici:1995am}. This idea lead 
to the development of 
the two-step 
multi-boson (TSMB) algorithm \cite{Montvay:1995ea}, where two polynomials are 
constructed, ${\cal P}_1(x)$ (low-order) and ${\cal P}_2(x)$ (high-order) such 
that 
\begin{equation}
  {\cal P}_1(x) {\cal P}_2(x) \approx \frac{1}{x^{N_f/2}}.
\end{equation}
${\cal P}_1(x)$ is replaced by a partition function of a small number of boson
fields, and ${\cal P}_2(x)$ is used in the acceptance test.

  Polynomial approximations have been incorporated into other
techniques. The Polynomial Hybrid Monte Carlo (PHMC) algorithm
\cite{Frezzotti:1997ym} proposed by Frezzotti and Jansen is one such 
development. A single auxiliary field, $\phi$ is introduced, rather than the 
large number of LBA, and the action is 
\begin{equation}
  S_\phi = S_G[U] + \phi^* {\cal P}(M^\dagger M) \phi
\end{equation}
This avoids the stiff update dynamics, however the action is now 
non-local, so HMC is used to update the gauge fields. 
The algorithm can be made exact either via an accept-reject step or by a 
re-weighting of expectation values by a correction factor:
\begin{equation}
  \langle {\cal O} \rangle_{\rm true} = \frac 
      {\langle{\cal O}\; \det Q^2 {\cal P}(Q^2)\rangle_\phi}
      {\langle         \det Q^2 {\cal P}(Q^2)\rangle_\phi}.
\end{equation}
Frezzotti and Jansen
advocate this re-weighting and the break-down of the approximation close to 
zero is now regarded as advantageous. 
The polynomial weight generates more configurations with low eigenvalues (which
are then assigned a lower weight in the ensemble average), and this 
over-sampling should lead to a more reliable determination of quantities that
are sensitive to the details of the lowest eigenvalues. 

The ALPHA collaboration have made a comparison \cite{Frezzotti:2000rp}
between PHMC and pseudofermionic HMC. In the range of 
parameters of interest, they find there is a slight advantage to PHMC and
favour this method for future simulations, on account of its over-sampling 
property.  Schroers presented results to the conference \cite{Schroers:2001if}
that suggest the performance of the tuned LBA is faster than the SESAM 
production HMC code.  Performance comparisons between TSMB and HMC in compact 
QED were presented to the conference by Zverev \cite{Bogolyubsky:2001hc}. The 
study finds the two algorithms are equally competitive, although the authors 
emphasise their TSMB implementation can still be optimised by changing the 
local gauge update scheme and tuning the polynomials ${\cal P}_1$ and 
${\cal P}_2$ more carefully.

One remarkable
feature of these comparisons is that the results are, in fact, so close. 
de Forcrand emphasises that the LBA algorithm turns into a standard local
heatbath/over-relaxation algorithm as the fermions are made heavier, and the
expectation there is that the over-relaxation algorithms should out-perform
(global) HMC significantly. 

\section{ODD-FLAVOUR SIMULATIONS}

  Since QCD has three flavours of light fermions, there is a natural 
incentive to study algorithms for simulating an odd number of fermion flavours. 
The standard pseudofermion formulation is unsuitable, 
since the Wilson matrix can have eigenvalues with negative
real parts, meaning there are regions of configuration space where the 
gaussian integral is not defined. 
  One extremely useful development arising from the local boson method has been
the possibility of performing these odd-flavour simulations,
even maintaining an exact update algorithm. 
Since these simulations still rely on gaussian integrals, they in fact generate
configurations with probability measure defined from the modified partition
function,
\begin{equation}
  Z_+ = \int\!\!{\cal D}U \;|\det M|\; e^{-S_G}.
  \label{eqn:Zplus}
\end{equation}
The sign of the determinant can be included in a {\it post-hoc} reweighting of
observables:
\begin{equation}
  \langle {\cal O} \rangle_{\rm true} = \frac 
      {\langle{\cal O} \mbox{ sgn}(\det M)\rangle_+}
      {\langle         \mbox{sgn}( \det M)\rangle_+}.
\end{equation}
This may lead to a ``sign problem'', where statistical estimates of reweighted
observables fluctuate wildly if a mixture of configurations with both positive 
and negative sign determinants occur in the ensemble. Empirical studies of this
problem for new simulations of QCD with three quarks (with a target of 
$m_q \approx m_s/4$) were presented to the conference by Gebert 
\cite{Farchioni:2001di}. They expect the sign problem to be very mild in this 
region.

The local boson algorithm can approximate the partition function of Eqn. 
\ref{eqn:Zplus}; starting from a polynomial approximation to $1/\sqrt{x}$ in 
the range $[\epsilon,1]$ leads directly to the one-flavour partition function.
Polynomials of the non-hermitian fermion matrix $M$ also present a solution, as
suggested in Ref. \cite{Borici:1995am} and developed in Ref. 
\cite{Takaishi:2001um,Aoki:2001yj}. A polynomial approximation to $M^{-1}$ of 
order $2n$ is constructed,
\begin{equation}
  M^{-1} \approx {\cal P}(M) = \prod_{k=1}^{2n} (M - z_k).
\end{equation}
For a suitably chosen polynomial, the roots, $z_k$ come in complex-conjugate 
pairs, so the polynomial can be written
\begin{equation}
  {\cal P}(M) = \prod_{k=1}^{n} (M - z_k) (M - z_k^*).
\end{equation}
The $\gamma_5$ hermiticity of $M$ means
\begin{equation}
  \det (M - z_k^*) = \det (M^\dagger - z_k^*)
\end{equation}
so $|\det M|$ can be represented approximately by a gaussian integral,
\begin{equation}
  |\det M| \approx \int\!\!{\cal D}\phi {\cal D}\phi^*
       \exp \left\{ -\phi^* T^\dagger_n(M) T_n(M) \phi \right\}
\end{equation}
with
\begin{equation}
  T_n(M) =  \prod_{k=1}^{n} (M - z_k).
\end{equation}
The scheme can be made exact with a Metropolis test; two alternatives
were proposed in Refs 
\cite{Takaishi:2001um,Aoki:2001yj}, both of which rely on a noisy 
Kennedy-Kuti \cite{Kennedy:pg} acceptance test. 
The three-flavour simulations of QCD are being carried out by the 
JLQCD collaboration and results were presented to the conference 
\cite{Aoki:2001xq}.

\section{ACCELERATING HMC}

HMC remains the most 
popular technique for generating ensembles of gauge fields with two flavours of 
dynamical Wilson (and Sheikholeslami-Wohlert improved) fermions. Most of the 
simulations use this well-established method, so it is useful to find simple
schemes to enhance the method.

\subsection{ILU preconditioning \label{sec:ilu}}

  Replacing the fermion determinant by a gaussian integral couples extra
stochastic degrees of freedom to the gauge fields. This in turn leads to a 
lower Metropolis acceptance rate in HMC. It has been known for a long time that
coupling the gauge fields and pseudofermions with the even-odd preconditioned 
fermion matrix (whose determinant is equal to that of the original fermion 
matrix) leads to a higher acceptance rate for a given molecular dynamics 
step-size  \cite{eo}.
The SESAM collaboration \cite{Fischer:1996th} emphasised that the 
even-odd preconditioning scheme is an example of an incomplete LU factorisation 
technique for matrix preconditioning. ILU preconditions a matrix by left and 
right multiplication by two readily invertible matrices;
\begin{equation}
  \tilde{M} = (I-L)^{-1} M (I-U)^{-1}
\end{equation}
where L and U are the lower and upper sections of the matrix. Defining L and U
first requires sites on the lattice, $\bf x$  are assigned an index, 
$s(\bf x)$ then site ordering is defined as 
\begin{equation}
  {\bf x} > {\bf y} \mbox{\hspace{4ex}if\hspace{4ex}} s({\bf x}) > s({\bf y}).
\end{equation}
SESAM developed a site ordering that is well-suited to accelerate matrix 
inversion on parallel computers. Since this preconditioning leaves the 
determinant of the matrix invariant, $\tilde{M}$ can
be used to couple the pseudofermions to the gauge fields. This holds for any
choice of ordering function, $s({\bf x})$. In Ref.  \cite{Peardon:2000si},
different site orderings 
were tested in HMC simulations of the two-flavour Schwinger model. An unusual
pattern emerged; the optimal ordering for speeding up matrix inversion was the
global lexicographic scheme, while the best ordering for
improving Metropolis acceptance was the usual even-odd preconditioner. In 
Ref. \cite{deForcrand:1997ck}, it was noted that the even-odd matrix itself can 
be ILU factorised again, and this both accelerates matrix inversion and 
improves the Metropolis acceptance rate. Ref. \cite{Peardon:2000si} tested a 
range of two step (eo-ILU) preconditioned matrices (defined now by different 
choices of an indexing function, $s({\bf x}_e)$ on the even sub-lattice only). 
These simulations found the same pattern emerged; the optimal ordering for 
inversion is a globally lexicographic one, while using a ``locally ordered'' 
scheme is best for improving Metropolis acceptance. Using an eo-ILU ordered 
interaction allows a three times larger step-size for a given Metropolis 
acceptance rate.  The naive expectation that the autocorrelations of the 
update scheme are independent of the integrator step-size at a fixed acceptance 
is supported by simulation data.
\begin{figure}
  \vspace{1ex}
  \setlength{\epsfxsize}{7cm} \epsfbox{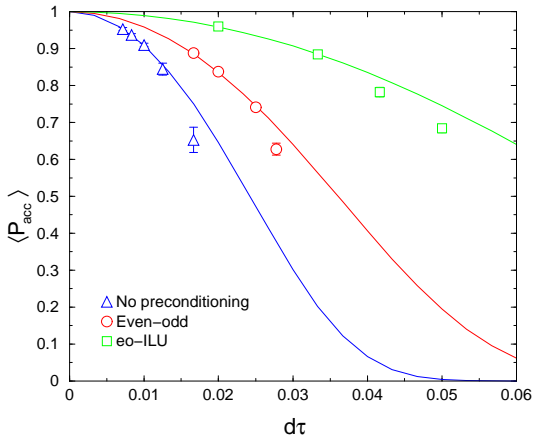}
  \vspace{-6ex}
  \caption{Schwinger model HMC Metropolis acceptance probability for 
pseudofermions coupled via different preconditioned matrices 
\protect{\cite{Peardon:2000si}}.}
\end{figure}

\subsection{Splitting the pseudofermions}

  Hasenbusch \cite{Hasenbusch:2001ne} recently proposed modifying the 
pseudofermion sector of HMC to improve the acceptance rate for a fixed 
molecular dynamics step-size.  The key observation is (as above) that coupling 
the gauge fields to pseudofermions via a well-conditioned matrix allows a 
larger molecular dynamics step-size to be taken for a fixed Metropolis 
acceptance. The fermion determinant is first split into two pieces, namely
\begin{equation}
  \det M = \det M \bar{M}^{-1} \det \bar{M}.
\end{equation}
The two-flavour determinant is then represented by a gaussian integral,
\begin{equation}
  \det M^2 = \int \!\!
    {\cal D}[\phi, \phi^* \psi, \psi^*]
      \exp \left\{ -S_\phi - S_\psi\right\}
\end{equation}
with
\begin{equation}
  S_\phi=| \bar{M} M^{-1} \phi | ^2 \mbox{\hspace{2ex}and\hspace{2ex}} 
  S_\psi=| \bar{M}^{-1}   \psi | ^2.
\end{equation}
Hasenbusch then chooses $\bar{M} = I - \tilde{\kappa} \Delta$ and recognises 
that for $0 < \tilde{\kappa} < \kappa$, the two matrices $ M \bar{M}^{-1}$ and 
$\bar{M}$ are better conditioned than $M$. This leads to improved 
acceptance. 
\begin{figure}[h]
  \vspace{-3ex}
  \setlength{\epsfxsize}{7.7cm} \epsfbox{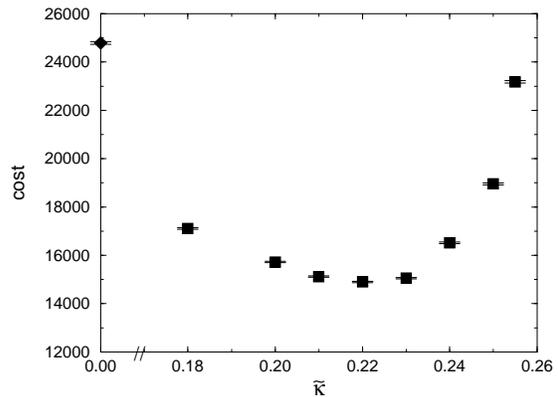}
  \caption{Schwinger model simulation cost vs $\tilde{\kappa}$ for the split 
pseudofermion scheme \label{fig:hasenbusch} 
\protect{\cite{Hasenbusch:2001ne,Hasenbusch:2001xh}}.}
\end{figure}
In Ref \cite{Hasenbusch:2001ne}, Hasenbusch tested the idea in the 
two-flavour Schwinger model and at the conference \cite{Hasenbusch:2001xh},
results for preliminary studies in QCD were presented. 
The cost of HMC simulations (defined in units of matrix $\times$ vector 
operations) as a function of the splitting parameter, $\tilde{\kappa}$ is 
summarised in Fig. \ref{fig:hasenbusch} In the Schwinger model 
study, simulations are accelerated by a factor of 1.7 when the pseudoscalar 
meson mass, $m_P = 0.210(3)$, rising to a speed-up of 2 when $m_P=0.124(5)$. 
This increase at lighter fermion mass is an encouraging result. The studies in 
QCD \cite{Hasenbusch:2001xh} are more exploratory, but effects of similar 
magnitude are seen. 

\subsection{Multiple time-scale integration \label{sec:integrator}}

  Based on the idea of separating the infra-red and ultra-violet sectors of 
the fermion matrix,
Jim Sexton and I are working on a 
practical implementation of a multiple time-scale molecular dynamics 
integrator. A low-order polynomial approximation to the fermion matrix inverse, 
which captures the short distance scale effects of quark loops is used, along 
with a pseudofermion action to reproduce long-range physics. The force on the 
gauge field from the polynomial term in the action is integrated (along with 
the easy-to-compute force from the Yang-Mills discretisation) using a finer 
time-step, and the long-range modes, contained in the pseudofermion action are 
integrated using a larger time step  \cite{Sexton:1992nu}. 

\subsection{Are these schemes compatible?}

  If these ideas could be combined, the possibility of an 
order-of-magnitude acceleration of Wilson fermion simulations would open up. 
  ILU preconditioned pseudofermions (Sec. \ref{sec:ilu}) can be used 
straightforwardly in the multiple time-scale scheme of Sec. 
\ref{sec:integrator}. This idea is under investigation, and it seems likely the
benefits from each component will be combined. Hasenbusch demonstrated directly
that his scheme works with even-odd preconditioned pseudofermions. 

\section{NEW DEVELOPMENTS}

The Kentucky group \cite{Lin:1999qu} are developing a new dynamical fermion 
updater based on a Kennedy-Kuti noisy acceptance test on changes in the 
determinant. The ratio of the determinants evaluated on the new and old 
configurations is computed using a $Z_2$ stochastic estimate of the trace of 
a Pad\'e approximation to the logarithm of the fermion matrix.
The scheme is still being developed. One interesting feature \cite{Joo} is the 
possibility of using the algorithm to simulate finite density QCD, by 
including a projection onto states with definite particle number. 

Bakeyev \cite{Bakeyev:2001wb} has proposed a new scheme for performing 
exact, finite updates of the gauge fields interacting with dynamical fermions.
The update involves finding a new configuration, $\tilde{U}$ that solves 
\begin{equation}
  M(\tilde{U}) \eta = \omega M(U) \eta
\end{equation}
for a stochastic parameter, $\omega$ and gaussian noise source, $\eta$. The 
idea is in its infancy as yet, and may prove to be most useful as an 
additional step in an HMC simulation. 

The flavour-symmetry breaking of staggered fermions can be reduced by using a
``fat link'' to induce quark-gluon interactions. Simulating these staggered
actions poses a new problem in algorithm design. The fat link is 
constructed following a similar procedure to APE smearing, which involves 
summing the six staples around a link, then projecting the resulting matrix 
back into the gauge group. The fattening step is repeated iteratively. The 
challenge for simulation is that the force term on the underlying gauge fields 
is difficult to compute. New algorithms are being developed and tested 
to circumvent this problem. Knechtli \cite{Hasenfratz:2001mi} presented these 
ideas to the conference. HMC fat-link simulations with four flavours when the 
lightest pion has mass $r_0 m_\pi=2.0$ are seven times more costly than runs 
where the quarks interact directly with the link variables. This factor is 
reduced to about 3 once $r_0 m_\pi$ reaches 1.6
This extra overhead is more than offset by 
the much better flavour symmetry of the fat-link actions, and the quark mass
accessible to these studies should be much lighter. 

\section{HIERARCHICAL NOISE REDUCTION IN YANG-MILLS THEORIES \label{sec:LW}}

\begin{figure}
  \vspace{-2ex}
  \setlength{\epsfxsize}{7.5cm} \epsfbox{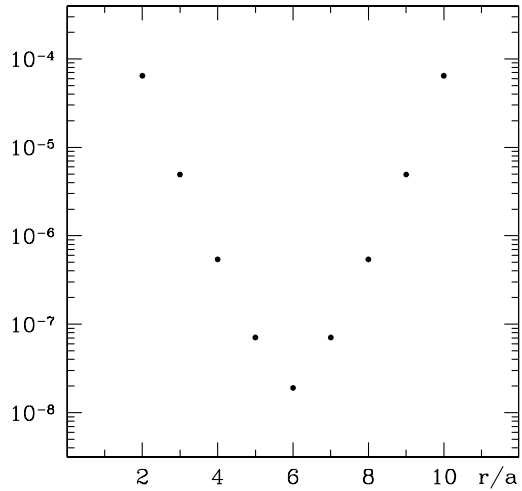}
  \vspace{-8ex}
  \caption{The Polyakov loop correlator on a $12^4, \beta=5.7$ lattice, 
computed using the hierarchical noise reduction method. Statistical errors are 
smaller than the plot symbols \protect{\cite{Luscher:2001up}}.\label{fig:lw}}
  \vspace{-3ex}
\end{figure}

  Recently, L\"uscher and Weisz \cite{Luscher:2001up} proposed a novel 
algorithm for reducing the 
stochastic noise in large Wilson loop expectation values by a factor that grows exponentially with the loop size. The scheme involves 
building a hierarchy of sub-domains within the lattice, and performing 
a reduced Monte-Carlo integration over the degrees of freedom inside each
domain. The results at each level are passed up the hierarchy to construct
variance reduced averages of larger and larger objects. The scheme is a 
natural extension of the temporal link ``multi-hit'' method 
\cite{Parisi:1983hm}. They illustrate their 
proposal with a calculation of the correlator of two spatially-separated 
Polyakov lines (see Fig. \ref{fig:lw}), and show a reduction of many orders of 
magnitude in the statistical error over the standard technique.

  The scheme can be extended to include matter fields. If the fermion 
determinant is included using the LBA, the action is 
localised within a small neighbourhood. With Wilson fermions (and the 
Wilson parameter set to unity) the hierarchical measurement scheme can be 
incorporated without modification. The difficulty is including the stochastic
accept-reject schemes usually employed with the LBA to make 
the method exact. These rely on lattice-wide observations, involving all the 
gauge degrees of freedom and the domain decompositions of Ref 
\cite{Luscher:2001up} would no longer apply. 

  It remains an unsolved challenge to include the highly developed tools for
optimising ground-state overlap into the hierarchy scheme. Methods such as
APE smearing and Teper blocking have proved crucial in the accurate 
calculations of the inter-quark potential and its excitations in Yang-Mills 
theory. Variational techniques, which build a large basis sets of creation
operators for the states of interest, then find an optimal ground-state 
operator by diagonalisation have also dramatically improved these calculations
(for an example, see Ref. \cite{Juge:2001mj}).
While the theoretical aspect of including these ideas is readily resolved, it 
is not obvious how well this toolkit can be used within the hierarchy method in practice.  An empirical test seems the only way to resolve this issue. If the 
noise reduction of the method of Ref. \cite{Luscher:2001up} can be married to 
the excellent
ground-state construction of the smoothing and operator basis methods, then 
extremely precise calculations of the detailed nature of the confining string
spectrum at large separations could be performed. 

\section{ALL-TO-ALL FERMION PROPAGATORS}

  Having expended such a considerable number of cycles on some of the world's 
largest supercomputers in generating an ensemble of gauge field configurations,
it is clear we are duty bound to make the most of the information they contain.

  For many observables, believed to be sensitive to vacuum fluctuations in the
quark fields, traditional point-source propagator methods are inefficient. 
To extract a useful signal from the ensemble, the propagator from all (or many)
points to all points on the lattice is needed. There has been a good deal of 
interest in improving these techniques in recent years. 

\subsection{Gaussian and $Z_2$ stochastic estimators}

  A gaussian representation of a fermion matrix entry can be written
\cite{Fucito:1980fh}:
\begin{equation}
  Q_{ij} = \int\! {\cal D}\phi {\cal D}\phi^* \;\; \phi_i (\phi^* Q)_j \;\exp
     \left\{ -\phi^* Q^2 \phi \right\}.
\end{equation}
Thus a propagator from any point on the lattice to any other can be estimated
by performing a sub-Monte Carlo simulation on each configuration.

  At the conference, Duncan \cite{Duncan:2001dv} presented simulation data 
using a gaussian 
all-to-all scheme. He emphasised that for many interesting physics applications,
this simplest of methods is adequate to ensure the dominant source of 
statistical error is from fluctuations within the gauge ensemble, rather than
the stochastic estimator. The stochastic degrees of freedom can be updated in 
a global heatbath step, but this requires a matrix inversion. Ref.
\cite{deForcrand:1998je} gives a scheme to accelerate the process; the 
iterative matrix solver is stopped after a small number of steps, and a 
Metropolis test is used to correct for the error in this step. 

  The Kentucky group \cite{Dong:1993pk} noted that other stochastic variables 
can be used to compute matrix traces, and exploited random elements of 
$Z_2$. These estimators have been employed in a range of calculations involving
disconnected diagrams.

\subsection{Improving stochastic estimators}

Ref. \cite{Michael:1998sg} suggests reducing the noise in a gaussian estimator 
by dividing the lattice into two segments, then
computing the independent sets of integrals in each segment exactly in a 
fixed background of the degrees of freedom on the interface between the 
segments. Conceptually, the new L\"uscher-Weisz 
hierarchy method, described in Sec. \ref{sec:LW} is reminiscent of this scheme. 
The method has one restriction; the source and sink points of
the propagator can not be in the same domain, so closed fermion loops can not
be computed. In Ref. \cite{McNeile:2000xx}
a further enhancement of the method is presented. 

At the conference, a method was presented by Wilcox \cite{Wilcox:2001qq} to mix 
gaussian and $Z_2$ noise, allowing an approximate heatbath with Metropolis 
correction to be used.

The SESAM collaboration \cite{Viehoff:1997wi} improved $Z_2$ estimators by 
breaking the vector space into sub-spaces (in their study, the 4 sub-spaces 
corresponding to the spin indices) and performing separate stochastic 
estimators in each sub-space. The idea was tested further in Ref 
\cite{Wilcox:1999ab}.
While this means more matrix inversion must be 
performed for a fixed stochastic ensemble size, the resulting estimators can 
have a significantly smaller variance. This idea of thinning can be extended 
beyond just spin and colour indices to spatial 
sites as well. If the scheme were to be carried to the extreme, where the 
$N$ dimensional vector space is decomposed into $N$ 1-dimensional spaces, a 
$Z_2$ estimator would reproduce the exact result for any ensemble size. This is
not such a surprising result, since this computation would require $N$ 
inversions to be performed and it is a 
trivial result that the trace of $M^{-1}$ can be computed in $N$ inversions.
Without thinning however the error falls off, as any statistical estimate 
should, like $1/\sqrt{N}$ so eventually thinning must become more 
efficient. 

  The thinning procedure has not been explored fully for computations of 
all-to-all propagators in QCD, and it would be worthwhile to investigate this
more completely.

\subsection{Eigenvalue decomposition}

  If all the eigenvectors (and eigenvalues) of the fermion matrix are known, 
then any propagator entry can be computed straightforwardly. It is natural to
expect that the long-range physics of QCD are contained in the lowest-lying 
eigenvectors, and these will then dominate in the spectral 
representation of a mesonic correlator (for example). The importance of 
understanding the physics of the lowest modes of the fermion matrix was 
emphasised in the plenary talk of Edwards \cite{Edwards}.

\begin{figure}
  \setlength{\epsfxsize}{7.7cm} \epsfbox{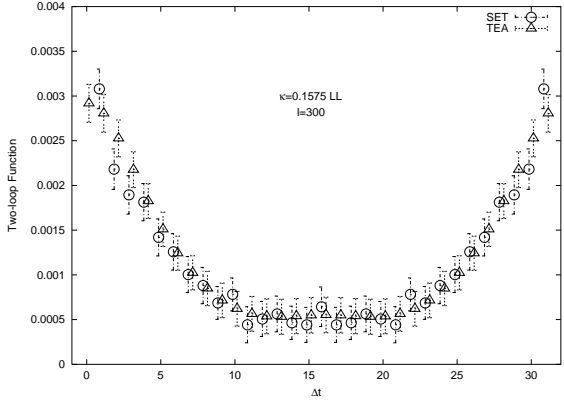}
  \vspace{-6ex}
  \caption{The disconnected (hairpin) correlator for the $\eta'$, computed 
using both an eigenvalue decomposition (TEA) and stochastic estimators (SET)
\protect{\cite{Schilling:2001xd}}.}
\vspace{-3ex}
\end{figure}

Neff presented the conference \cite{Neff:2001xc} with an investigation of 
methods for computing the low-lying eigenvectors efficiently. The acceleration 
comes from using a polynomial to isolate the region of eigenvalues of interest,
and this leads to more rapid converges of the Arnoldi method. 
At the conference, Schilling \cite{Schilling:2001xd,Neff:2001zr}
discussed the dominance of the $\pi$ and $\eta$ correlators by the low-lying
eigenmodes. 
Truncating the spectral expansion introduces a systematic error, but this can 
be avoided by computing all-to-all propagators in a hybrid scheme (as noted by
Edwards \cite{Edwards}).

\subsection{Hybrid schemes}

  The two schemes can be combined straightforwardly to construct a hybrid 
all-to-all estimator method \cite{Edwards}. 
A number of low-lying eigenvalues of the fermion
matrix are computed and incorporated exactly and 
the effect of all the remaining eigenvalues is estimated
stochastically. This involves building a projection operator onto the reduced
sub-space spanned by the explicitly computed eigenvectors. 

For illustration, consider the observable
\begin{equation}
  \mbox{Tr } Q^{-1} \Gamma =
     \sum_i^N \frac{1}{\lambda_i} \;\;v_i^*\; \Gamma v_i^{}.
\end{equation}
If the lowest $m$ ($m \ll N$) eigenvectors of the hermitian matrix 
$Q=\gamma_5 M$ are 
computed, then it is natural break the 
the spectral representation of $Q$ into two parts,
\begin{equation}
  Q = \sum_{i=1}^m \lambda_i^{} v_i^{} v_i^* + 
      \sum_{i=m+1}^N \lambda_i^{} v_i^{} v_i^*.
\end{equation}
Defining the (sub-space) inverses and projectors,
\begin{equation}
\bar{Q}_{(0)} = 
   \sum_{i=1}^m \frac{1}{\lambda_i}\;\; v_i^{}\;v_i^*
\mbox{,\hspace{3ex}}{\cal P}_{(0)} = 
   \sum_{i=1}^m \; v_i^{}\;v_i^{\textstyle *}
\end{equation}
the trace becomes
\[
  \mbox{Tr } Q^{-1} \Gamma = \mbox{Tr } \bar{Q}_{(0)} \Gamma +
                                  \mbox{Tr } \bar{Q}_{(1)} \Gamma.
\]
The first term is the truncated estimate, and the remaining term can be 
estimated stochastically;
\begin{eqnarray}
    \mbox{Tr } \bar{Q}_{(1)} \Gamma & = &
       \langle \eta^{\textstyle *} \Gamma \bar{Q}_{(1)} \eta \rangle
              \nonumber\\
                                         & = &
       \langle \eta^{\textstyle *} \Gamma (Q^{-1} - \bar{Q}_{(0)})
           \eta \rangle
              \nonumber\\
                                         & = &
       \langle \eta^{\textstyle *} \Gamma Q^{-1} (I  - {\cal P}_{(0)})
           \eta \rangle.
\end{eqnarray}
The operation of $I  - {\cal P}_{(0)}$ is an orthogonalisation of the noise 
vector, $\eta$ with respect to all $m$ known eigenvectors. The matrix inversion
in this last step is accelerated, since the eigenvectors corresponding to the 
low-lying eigenvalues have been removed from $\eta$. 
It will be interesting to see hybrid schemes tested in practical applications.
The developments of GMRES algorithms presented to the conference 
\cite{Morgan:2001qr} look particularly interesting in this context. 

\section{CONCLUSIONS}

  For the next few years, most large-scale simulations of full QCD will use the
Wilson or staggered fermion formulation, while the chiral lattice actions 
\cite{Hernandez:2001yd} are beyond the reach of current computing resources; 
Ginsparg-Wilson fermions are a factor of 100 times more expensive.  
As pointed out by Karl Jansen, the Hybrid Monte-Carlo 
algorithm is not far off its twentieth birthday and remains (largely 
unmodified) the most popular tool for dynamical Wilson fermion simulations. 

Current predictions \cite{rdkpanel} are that resources available in the next
few years are an order of magnitude below the requirements for reliable QCD 
simulations. Including the newest ideas presented, it might be hoped to
get close to the required factor of ten improvement in performance. 

  Realistic comparative tests of HMC vs LBA have been carried out in recent 
years, leading to the observation that the two methods have identical costs 
(for all practical purposes). These comparisons are always very difficult to 
make. It would be helpful to see a standard benchmark emerge on which direct 
comparisons of algorithms can be made.  
Defining a standard is difficult but with this in place, fair races of new 
methods against ``thoroughbred'' algorithms rather than the simplest 
implementations could be made. 

  Excitement at the conference generated by the recent idea of L\"uscher and 
Weisz suggests reconsidering the traditional Monte-Carlo analysis pathway; 
generating configurations then subsequently making measurements. Their 
hierarchy scheme, which knits these two processes together can lead to orders 
of magnitude increases in efficiency. The challenge is to apply this
philosophy to a wider range of applications. 

  I am grateful to Ph. de Forcrand, C. Gebert, M. Hasenbusch, 
B. Jo\'o, A. Kennedy, J. Kuti, M. L\"uscher, I. Montvay, H. Neff, W. Schroers,
J. Sexton and N. Zverev for many stimulating discussions, observations 
and helpful correspondence.

\end{document}